\documentclass[aps,prl,preprint,groupedaddress]{revtex4}

\begin{document}


\title{General relativity, Type Ia supernovae and the accelerated expansion of the universe}


\author{Patrick Das Gupta}
\email[]{patrick@srb.org.in}
\affiliation{Department of Physics and Astrophysics, University of Delhi, Delhi - 110 007 (India)}


\date{\today}
\begin{abstract}
The 2011 Nobel prize in physics has been awarded to S. Perlmutter, A. Riess and B. Schmidt for their path breaking discovery that
the rate of expansion of the universe is increasing with time. The trio used Type Ia supernovae (SNe Ia) as standard candles to estimate their luminosity-distances. 
To appreciate some of the far reaching implications of their work, I have provided an elementary exposition of general theory of relativity, accelerated expansion of
 the universe, luminosity-distance, SNe Ia and the  cosmological constant problem.

\end{abstract} 

\maketitle

\section{Introduction}
 Gravity is always attractive, right? Wrong, if we are to go by the research papers of  S. Perlmutter, A. Riess and B. Schmidt, for which they  
have been awarded this year's Nobel prize in physics [1-5]. The trio looked at Type Ia supernovae, located in far away galaxies,
 to measure their distances accurately, and landed up with a surprising discovery -  cosmic anti-gravity at large scales.  To appreciate some
of the implications of their path breaking findings, and to have a glimpse of the repulsive side of gravitation, we need to look into
 gravity's fatal attraction, first.
   
 According to the Newtonian theory,  acceleration of a test body  due to a massive object's gravity, while being proportional to latter's inertial mass and 
  directed towards the massive body,  is
  independent of the test particle's inertial mass. Newton's gravity also demands that the gravitational
 force be instantaneously transmitted by the source to the test particle, since it is inversely proportional to the square of the  instantaneous separation between the two.
 Instant transmission is unsatisfactory, as Einstein's special theory of relativity demands that no physical effect can  propagate with a speed faster than $c$ (speed of light in vacuum).
 
     Einstein improved the situation by putting forward a relativistic theory of gravity in 1916 through his theory of general relativity (GR). GR is 
 based on the observation that the trajectory of a test particle in any arbitrary gravitational field is independent of its inertial mass $ m$ (since the acceleration does not dependent on $m$),
 and therefore, it must be the geometry of the space-time that determines test particle trajectories. Note that for no other force, acceleration of a test particle
 is independent of its inertial mass (e.g. in classical electrodynamics, acceleration of a test charge is proportional to the ratio of its charge to mass).

Imagine that a small bundle of test bodies are freely falling in an arbitrary gravitational field. Since their accelerations due to gravity are nearly identical, if one were to sit on one
 such particle and observe the rest, one would find that the other test particles are freely floating as though gravity has simply disappeared! This is the  principle of equivalence which
states that no matter how strong or how time varying the gravity is, one can always choose a small enough frame of reference for a sufficiently small time interval such that gravity vanishes
in this frame. 

This small region is a locally inertial frame of reference, and laws of physics in this frame
 take the same form as they do in special theory of relativity. In special relativity, the proper distance $ds $ between two nearby events with space-time coordinates
 $x^\mu$ = $(ct,x,y,z)$ and $x^\mu + dx^\mu$ = $(ct +cdt,x + dx, y+dy, z+dz)$
 is given by,

 $$ds^2 = c^2 dt^2 - dx^2 - dy^2 - dz^2 \equiv \eta_{\mu \nu}dx^\mu dx^\nu \ . \eqno(1)$$
Note that in eq.(1),  $x^i$, i=1,2,3 are the  Cartesian coordinates of the event, and $\eta_{\mu \nu} $ is the Minkowski metric with $\eta_{00}=1= -\eta_{ii}$, rest of the components of the metric 
 being zero.
  Einstein's summation convention has been used in eq.(1), so that repetition of Greek indices imply summation over 0,1,2 and 3. 

In 3-dimensional Euclidean geometry, the line-element $dl^2= dx^2+dy^2 + dz^2 $ has
the same form whether you shift the Cartesian coordinate system by any constant vector or rotate the coordinate system about any axis by any constant angle. The line-element given by eq.(1) is
 similarly invariant under  Lorentz transformations as well as constant space-time translations. According to the equivalence principle, whatever is the  gravity around, in a locally inertial
 frame (i.e. freely falling frame), the line-element is given by eq.(1) and non-gravitational laws of physics take the same form as in special relativity. 
  But, what is the connection between this feature of gravitation and geometry?

 Consider a  generally curved two-dimensional surface (e.g. the surface of, say, a pear). No matter how greatly the surface is curved, one can always choose a tiny 
enough patch on it, such that it is
 sufficiently flat for Euclidean geometry to hold good over it. As one increases the size of the patch, the curvature of the pear's surface  becomes apparent.  This is
 so similar to the main characteristic of gravity  that  we discussed in the
 preceding paragraph. The small patch on the pear over which the line-element is Euclidean ($dl^2= dx^2 + dy^2 $) is analogous to the local inertial frame in the case of 4-dimensional space-time
 where the line-element is described by eq.(1).

 By going over to a small freely falling frame and choosing inertial (many a times called Minkowskian) coordinates, one manages to make gravity vanish so that  special relativity
  is all that one needs to describe laws of physics, locally. What if one wants to study laws of physics over larger regions of space-time? In that case, one would have to employ other coordinates
  that are curvilinear in general. Then it ensues that, instead of the  Minkowski metric, one would  require a general metric tensor.

 The fundamental entity in GR that describes space-time geometry is the space-time dependent metric tensor $g_{\mu \nu}(x^\alpha)$,  which  determines the invariant proper distance $ds$
 between any two nearby events with coordinates $x^\mu$ and $x^\mu + dx^\mu $,

 $$ds^2 = g_{\mu \nu}dx^\mu dx^\nu \eqno(2)$$
 Here, $x^\mu $, $\mu, \nu$=0,1,2,3, now represents a general curvilinear coordinate, specifying the location of an event. The metric $g_{\mu \nu} $ is a generalization of  $\eta_{\mu \nu}$,
 the Minkowski metric tensor. If the space-time geometry was not curved, one could choose a coordinate system such 
that everywhere the metric tensor is just the Minkowski metric tensor. But GR states that energy and momentum associated with matter warp the space-time geometry, entailing that in general it is
not possible to choose inertial coordinates everywhere so that the metric is globally Minkowskian.

 However, according to the principle of equivalence, by choosing an appropriate coordinate system,
 even in an
arbitrarily curved space-time, the metric tensor can be made to take the form of $\eta_{\mu \nu}$ in a sufficiently small space-time region (physically, this corresponds to choosing a sufficiently
small freely falling frame). In GR, the mathematical forms of physical laws remain the same even when one makes an arbitrary coordinate transformation. 

Although in a local inertial frame, gravitational force disappears, tidal force does not. For instance, earth is freely falling towards the sun because of latter's pull. But we do not feel sun's
gravity since the freely falling earth constitutes a local inertial frame. However, as sun's gravity is non-uniform, portions of earth closer to the sun feel a greater tug
 than those  located farther. This differential pull is the source of tidal force which causes the commonly observed ocean tides. In GR, the tidal acceleration is due to a fourth rank
 tensor called Riemann tensor that is
constructed out of the metric and its first as well as second derivatives. Therefore, the ocean tides owe their existence  to the non-zero Riemann tensor describing the 
  curvature of space-time geometry around the sun (as well as the moon).

  GR tells us that matter distorts the space-time from an Minkowskian geometry to a non-Minkowskian one, and test bodies just move along
 straightest possible paths in such a curved space-time. As to, how the matter warps the space-time geometry, is given by the so called Einstein equations
 which relate tensors created out of the metric and the Riemann tensor to the matter energy-momentum tensor multiplied by a combination of Newton's constant G and light speed c.
 Einstein equations possess a pristine beauty, with space-time geometry on one side, and the energy and momentum of matter on the other.
 When the geometrical curvature of space-time is small and the motion within the source  is slow enough, GR leads automatically to Newton's laws of gravitation.
 
   GR as a theory of gravitation gained  immediate acceptance among the physics community as soon as its prediction of bending of light was actually seen
 during the solar eclipse of 1919. Of course, GR had already correctly explained the anomalous precession of the perihelion of Mercury. Later, existence of gravitational waves (ripples
in space-time geometry) predicted by 
Einstein was also verified with the discovery of inspiralling Hulse-Taylor binary pulsar 1913 + 16. Indeed, gravitational effects too propagate as gravitational waves with finite speed c,
  consistent with the demands of  special relativity.

\section {Geometry of the Universe}

       Armed with these successes, Einstein in 1917 turned his attention towards building a general relativistic model of the whole universe. He found that  GR is unable to produce a static
 universe. Astronomers, those days, believed that stars and nebulae do not exhibit any large scale ordered motion. The prevalent view in the past  was that universe on large scales is static.
 GR, being a theory of attractive gravity, predicted that a large mass (like our 
universe) will either keep collapsing under its own weight or exhibit deceleration in its expansion rate if it was growing in size to begin with. In either case, a static universe from GR 
was out of question. Einstein
 was in a fix.

  His next step was to modify GR by introducing a term representing a kind of universal repulsion. This extra feature $\Lambda\ g_{\mu \nu}$, called the  cosmological constant term, on the left 
 hand side of Einstein 
 equations  helps in preventing gravitational collapse of the universe and, hence, leads to a static solution provided that value of the constant is positive (corresponding to cosmic repulsion).

 In 1929, the renowned American astronomer Edwin Hubble while studying the spectra of radiation from galaxies, discovered that spectroscopic lines were shifted to the red end as though
there was some kind of a Doppler redshift. The so called Hubble's law stated that the redshift $z$ of a galaxy satisfies $z= \frac {H_0} {c} d$, where $d$ and $H_0$= 72 km/s/Mpc are the
 distance of the galaxy from us and
 Hubble constant,
 respectively.
 Therefore,  it appears that galaxies recede from each other with a speed proportional to their distance of separation. According to Hubble's law, if the distance between two galaxies is 100 Mpc (i.e. about 326 million light years), the separation between the galaxies
 would be increasing at the rate of about 7200 km/sec. The universe actually is not static at all! Instead, the universe is expanding.
   Einstein called the action of introducing a cosmological constant term in GR to be his greatest blunder. Why? Because, he could have predicted that universe is not static, purely
 from theoretical calculations using his original theory. 

 In fact, way back in 1922, Friedmann had discovered (without using the  $\Lambda$ -term) 
 general relativistic solutions representing homogeneous, isotropic and expanding (or, contracting) universe. From the point of view of observations,
 although matter is clumpy and  inhomogeneous
 at small scales, if one considers distribution of galaxies and clusters of galaxies on scales larger than about 500 Mpc, the distribution of matter is remarkably homogeneous and isotropic.
Exploiting these symmetries, relativists like Friedmann, Lemaitre, Robertson and Walker could write down the line-element describing the geometry of such an isotropic and homogeneous universe,
$$ds^2=c^2 dt^2 - a^2(t) \bigg [\frac {dr^2}{1- k r^2} + r^2 (d\theta^2 + \sin^2 \theta \ d\phi^2) \bigg ]\eqno(3)$$
where the constant $k$ can take either the value 0 (spatially flat universe, infinite in extent), +1 (closed universe without a boundary, having a finite spatial volume) or -1 (open universe, infinite
in extent). 

It is evident from eq.(3), that the time dependent function $a(t)$ (called the scale factor) determines the spatial distance between a galaxy at $(r,\theta,\phi)$ and another at $(r+dr, \theta + d\theta,\phi + d\phi)$. An
increasing $a(t)$ with time describes an expanding universe. Hubble's law can be explained as follows: Light from a galaxy starts at time $t$ and after travelling an enormous distance, reaches us at
 time
$t_0$. During this period, the scale factor increases from $a(t)$ to $a(t_0)$, stretching the wavelength of light, and hence the redshift. The observed redshift is mainly due to the expansion of the universe and not due to any Doppler
shift. One can show that the redshift $z$ is given by,
$$\frac {\mbox{Observed wavelength}} {\mbox{Emitted wavelength}}\equiv 1 + \ z= \frac {a(t_0)}{a(t)}\eqno(4)$$
Farther the galaxy, earlier is the time $t$ of emission, smaller is $a(t)$, and therefore, larger is the redshift $z$.
 Distant galaxies and quasars have been observed with redshifts as high as 6, implying that radiation from such distant objects started when the universe was smaller by a factor of 7. The Hubble constant
is nothing but,
$$H_0 = \frac {1} {a(t_0)} \frac {da} {dt} \bigg \vert_0 \eqno(5)$$
The geometry described by eq.(3) is applicable only at cosmic scales (since homogeneity and isotropy is valid only at such large scales). 
 The sizes of planets, stars or galaxies do not get affected by the cosmological expansion.

If one takes the Friedmann-Robertson-Walker metric from eq.(3) and uses it in the Einstein equations corresponding to a homogeneously and isotropically 
distributed matter,  then one finds that the scale factor $a(t)$ satisfies,
$$\frac {{\dot {a}}^2 + k c^2} {a^2} =  \frac {8\pi G} {3 c^2} \epsilon \eqno(6)$$
and,
$$\ddot {a}= - \frac {4 \pi G} {3 c^2} \bigg (\epsilon + 3 p \bigg ) a \eqno(7)$$
where $\epsilon$ and $p$ are the total energy density and total pressure, respectively, of the contents of the universe.

When $k=0$ (flat universe) and matter is non-relativistic so that $p \ll \epsilon $, one finds from eqs.(6) and (7) that $a(t)$ is proportional to $t^{2/3}$.
 From eq.(4), one has the redshift-age relation for k=0 matter-dominated model,
$$ 1+z= \bigg ( \frac {t_0} {t} \bigg )^{2/3} \eqno(8)$$
For such a model, the Hubble constant (eq.(5)) is simply,
$$H_0= \frac {2} {3 t_0} \eqno(9)$$
 whereby the knowledge of $H_0$= 72 km/s/Mpc implies that the universe right now is about 13 billion years old.

From Friedmann-Robertson-Walker  equations (FRWE) (6) and (7), it follows that if the universe is just made up of standard matter like protons, neutrons, electrons, photons, neutrinos etc., the
 rate of expansion of the universe 
$\dot a$ slows
 down steadily with time ($\ddot {a} < 0$), as energy density and pressure are positive quantities. This is natural since gravity is attractive.
 Just to give you an analogy, when one
throws a ball up with a high velocity, although the distance between earth and the ball increases with time, the latter's velocity keeps decreasing because of earth's gravity. 

Had there been an
anti-gravity between the ball and earth, the velocity would have increased with time. Fortunately, inertial masses are always positive (otherwise, kinetic energy will be negative, leading to
 runaway situations, e.g. extraction of energy from a moving negative mass object would make the object travel faster!), entailing an attractive gravitation that keeps us grounded 
 to the earth.  But, is there no
possibility of gravitational repulsion?

\section {Repulsive gravity, luminosity-distance, Type Ia Supernovae and dark energy}

In GR, even  pressure associated with matter causes space-time geometry to be curved (e.g. eq.(7)). This is expected as pressure is related to the density of random kinetic energy
 of particles making up normal matter. Energy density is usually taken to be positive, otherwise matter would not be stable (as matter would keep decaying to more and
more negative energy density states). For standard matter, thermodynamic pressure is positive. Under such conditions, gravity can only be attractive.

 On the other hand, if a weird form of matter with a sufficiently large negative pressure permeates space, it will result in anti-gravity, since from eq.(7) the expansion rate would increase
 with time
(i.e. $\ddot {a} > 0$). From the first law of thermodynamics one knows that when ordinary gas  expands adiabatically, its internal energy decreases  since it performs work against the
 surroundings while growing in volume. Now imagine that one has expanding exotic matter (having negative pressure) instead. Its internal energy would then grow while expanding! 
 It is such counter-intuitive properties that lead 
to cosmic repulsion when GR is combined with presence of exotic matter.

For the cosmological constant, 
 one can show from the first law of thermodynamics that  while expanding, its internal energy grows in such way 
as to maintain a constant energy density (see eq.(14) below). It is of course consistent with the fact that cosmological constant is actually a constant .

As Einstein and Lemaitre had envisaged, a sufficiently large and positive  cosmological constant would lead to repulsive gravity on cosmic scales, since pressure $p_\Lambda $ in the case of 
cosmological constant $\Lambda$ is exactly
negative of its energy density $\epsilon_\Lambda $. 
 But how would one determine observationally whether in our universe the second derivative of $a(t)$ is positive or negative? This is where the investigations of Perlmutter, Riess and Schmidt assume 
importance. 
  
In the framework of Euclidean geometry, consider a source placed at a distance $d$ from us, emitting radiation isotropically with a luminosity
L. The flux of radiation $F$ received by us is
given by,
$$F= \frac {L} {4 \pi d^2}\eqno(10)$$
since  in unit time, $L$ amount of energy crosses a spherical
surface of area $4\pi d^2$. But when the
space-time geometry is curved as given by eq.(3), it can be shown that,
$$ F= \frac {L} {4 \pi d^2_L}\eqno(11)$$
where $d_L$ (defined to be the luminosity-distance) is given by,
$$d_L= a(t)(1+z)^2 r = a(t_0) (1+z) r\eqno(12)$$
where $r $ and $z$ are the radial coordinate and the redshift of the
source, respectively, while $t$ is the time when the radiation left the
source to reach us (i.e. $r=0$) at time $t_0$.

 In eq.(12), presence of  $a(t_0) r$ is normal since it is the physical distance of the source at the present epoch $t_0$. The factor $1+z$ arises 
because of the stretching of time intervals due to cosmological expansion. Photons radiated from the source in a time interval $dt$ is received by us over a period of $(1+z)dt $.  

Now, particles with zero rest mass like photons (light quanta) travel with speed $c$ in local inertial
frames, so that from eq.(1) one has $ds^2=0$ for light.
From equivalence principle, therefore, $ds^2= g_{\mu \nu}dx^\mu dx^\nu = 0$ for zero rest mass
particles in any coordinate system. In particular, in the cosmological setting, when light from a distant 
source moves along a radial $ds^2=0$ trajectory to reach us, one has from eq.(3), after taking the necessary square roots,
$$\frac {c dt} {a(t)}= - \frac {dr} {\sqrt {1-kr^2}}\eqno(13)$$

 One can solve the FRWE (eqs. (6) and (7)) to obtain $a(t)$ and thereafter, eq.(13) can be integrated to express the radial coordinate of the
source $r$ as a function of its redshift $z$, so that the luminosity-distance $d_L$ (eq.(12)) becomes a function of $z$, $H_0$ and the value
of $\ddot {a}$ at the present epoch $t_0$ [8]. 

Let us derive the expression for  luminosity-distance when $k=0$ and the cosmological constant
is positive, in an otherwise empty universe.
The energy density $\epsilon_\Lambda $ associated with a positive
cosmological constant $\Lambda $ is given by
$$\epsilon_\Lambda = \frac {c^4 \Lambda} {8 \pi G}\eqno(14)$$
with pressure,
$$p_\Lambda = -\epsilon_\Lambda .\eqno(15)$$
When eqs. (14) and (15) are substituted in the FRWE for $k=0$
case, one obtains,
$$\frac {\dot{a}^2} {a^2}= \frac {1}{3} \Lambda c^2 = H^2_0 \eqno(16)$$
$$\frac {\ddot{a}} {a}= \frac {1}{3} \Lambda c^2 \eqno(17)$$
implying that indeed there is cosmic repulsion, since $\ddot {a} > 0$.
In the flat and empty universe with a positive cosmological constant,
$\frac {\dot{a}} {a}$ does not change with time (eq.(16)), so that the
Hubble constant (eq.(5)) is independent of the epoch.

Solving the simple differential eq.(16) for an expanding universe, one obtains the scale factor,
$$a(t)= \exp {(H_0 t)} .\eqno(18)$$
Substituting eq.(18) in eq.(13) with $k=0$, one gets after integrating,
$$r=\frac {c} {H_0} [\exp {(-H_0 t)} - \exp {(-H_0 t_0)}]= \frac {c} {H_0}
\frac {z}{a(t_0)}\eqno(19)$$
where use has been made of the redshift-scale factor relation given by eq.(4).

 From eqs.(12) and (19), the luminosity-distance  in this model is simply,
$$d^{(\Lambda)}_L (z) = \frac {c} {H_0} z (1+z) \eqno(20)$$

Using a similar procedure, and making use of eqs. (8) and (13) with $k=0$, it is easy to work out
the following expression for
luminosity-distance in the case of flat matter-dominated model,
$$d^{(M)}_L(z)=\frac {2 c} {H_0} (1+z) \bigg [1 - \frac{1}{\sqrt{1+z}} \bigg ]
\eqno(21)$$

It is evident from a comparison of eqs.(20) and (21) that the luminosity-distance in the case of an accelerating universe increases with redshift much more rapidly than
that in a decelerating universe. In general, when there is matter as well as a non-zero cosmological constant, one can derive the expression for $d_L(z)$ following the above
procedure. Also, for $z \ll 1$, eqs.(20) and (21) both lead to the Hubble's law $d_L \approx \frac {c} {H_0} z$, telling us that the latter is only an approximate law, valid for 
smaller values of redshifts.

  One can show from FRWE that if the universe is permeated with a sufficiently large amount of strange kind of matter with an equation of state given by pressure $p_{DE}= \omega \epsilon_{DE}$ 
 with
 $  -1 \leq \omega \leq - \frac {1}{3}$, one has 
   $\ddot {a} > 0$ and luminosity-distance increasing with redshift at a faster rate than what happens in a decelerating universe. 
  This all pervading, exotic matter which does not emit or absorb light is generically referred to as dark energy (DE).
 The cosmological constant is a special
case of DE with $\omega = -1$. 

For an extragalactic source, the redshift as well as flux of radiation can be measured provided one has a sensitive telescope with proper spectro-photometric paraphernalia. So, 
if there exists a class of luminous sources which emit
radiation with almost a constant luminosity then by measuring their
redshifts along with fluxes, one can observationally determine their
 luminosity-distances (eqs.(11) and (12)) as a function of their redshifts. It so happens that 
 Type Ia supernovae (SNe Ia) are precisely what the doctor ordered. 

Supernovae basically are explosive events associated with dying stars. Types Ib, Ic and II supernovae are associated with very massive and evolved stars in which their iron core collapses, releasing
 vast quantity of energy on time scales of few milli-seconds, leading to violent explosions [6].  
On the other hand, a SNe Ia  explosion is driven by matter falling on a white dwarf star having initially a mass less than the Chandrasekhar limit of 1.4 $M_\odot $. Type II supernovae 
exhibit hydrogen lines, while supernovae of Types Ia, Ib and Ic are devoid of hydrogen lines in their spectra.
 
 A white dwarf, with a companion star, both going round their common centre of mass due to their mutual gravitational attraction, is likely to accrete matter steadily from its companion.
 At some point of time, the infalling matter raises the white dwarf's mass to above the threshold of Chandrasekhar limit.
 The core then becomes gravitationally unstable and implodes, giving birth to a SNe Ia.

 The sudden decrease in the
gravitational potential energy as the core collapses rapidly to a smaller
radius  results in the release of a huge amount of energy that blows
apart the star. A core with mass $M_c$, shrinking from a large size $R$
to a compact radius $R_c$, has to give up an energy,

$$E \sim \frac {G M^2_c} {R_c} \ \ ,\eqno(22)$$

since its gravitational potential energy decreases from $\sim -\frac
{G M^2_c} {R}$ to $\sim - E $ given by eq.(22). For a 1.4 $M_\odot $ core
collapsing to form a neutron star of radius $R_c \approx $ 10 km,
the explosion energy $E$ could be as high as $\sim 10^{53}$
ergs.

 The light curve of a typical supernova exhibits radiation luminosity rising with time rapidly to reach a peak value, followed by a gradual decay over a period of 30 to 40 days. 
  A large majority of SNe Ia have almost identical light curves and spectra, with a peak luminosity at visible wavelengths of about $2 \times 10^{43}$ erg/s. For the remaining SNe Ia,
  light curves are remarkably similar to each other displaying a nested character, enabling one to determine their intrinsic peak luminosities
  from the observed time scales of decay. In short, SNe Ia are good standard candles. 

Then, since $L$ is known, one
 can estimate $d_L(z)$ using eq.(11), after measuring flux $F$ and redshift $z$. These features were used to the hilt by Perlmutter, Riess and Schmidt. They discovered that at large redshifts the 
 observed extragalactic SNe Ia are significantly fainter than what is expected if the luminosity-distance is given by eq.(21) implying that indeed  $\ddot {a} > 0$ [7].

In Figure 1, redshift and peak flux, measured in units of $10^{-15}$ erg/s/cm$^2$, corresponding to 14 Type Ia supernovae (selected from reference [5]) have been plotted. 
The flux of supernova SN1992bl with redshift 0.043 is about $3.7 \times 10^{-12} $ erg/s/cm$^2$, while a more distant supernova SN1997G having redshift 0.763 is 
significantly fainter with a flux of about only $4.6 \times 10^{-15}$ erg/s/cm$^2$. If one assumes that these supernovae are standard candles with peak luminosity
$2 \times 10^{43}$ erg/s, the luminosity-distances given by eq.(11) are about 218 Mpc and $6.2 \times 10^3$ Mpc for SN1992bl and SN1997G, respectively. 

The upper and lower solid lines in fig.1 represent the theoretical flux-redshift relations (eq.(11)) corresponding to luminosity-distances $d^{(M)}_L(z)$ and
 $d^{(\Lambda)}_L(z)$, respectively (eqs.(20) and (21)), for a 
standard candle of luminosity $2 \times 10^{43}$ erg/s. It is evident from the figure that the $k=0$ matter dominated FRW model does not provide a good fit to the
 data at
  larger redshifts, as the observed supernovae are significantly fainter. 

On the other hand, the lower solid line corresponding to $k=0$ empty universe with a 
 positive cosmological constant predicts even fainter SNe Ia than observed at such high redshifts. Perlmutter, Riess, Schmidt and their collaborators showed 
 that a $k=0$ FRW model with 30 percent energy density in non-relativistic matter and 70 percent energy density lying with DE provides the best fit to the observed data.

Their results get support from current research studies which demonstrate that at the present epoch, only about 
30 percent  of the  total content of the universe is made up of ordinary matter plus weakly interacting dark matter, while roughly 70 percent rests with the DE component.
Cosmological constant is the most likely DE candidate, since so far all observations suggest that the equation of state parameter $\omega $ is very, very close to -1. But then, there is a fine tuning problem 
 associated with the cosmological constant. This riddle is still haunting
  fundamental physics.
  
\section{The $\Lambda $ Puzzle}

 From the point of view of classical GR, the only extra term that can be added to the Einstein equations (consistent with the local conservation of energy and momentum of matter) is a
 cosmological constant term. The physical dimension of this constant  $\Lambda $ is square of the inverse length. Observations tell us that the value of the cosmological constant, if not zero,
 is very 
small such that the physical size represented by  $\Lambda^{-1/2}$   is more than about 1000 million lightyears!
Hence, this repulsive aspect of gravity, if it exists, can only be felt at such large scales. Locally, gravity is essentially attractive.
    Why is the value of   $\Lambda $  so tiny? This is one of the burning questions of fundamental physics today. 
  
  According to quantum mechanics,  the ground state energy of a harmonic oscillator is non-zero. It is also a common knowledge that every physical field is associated with a 
particle. For example, electromagnetic field is associated with photons, Dirac field with electrons or quarks, and so on. There is some kind of a field-particle duality in quantum theory
  which is a generalization of de Broglie's wave-particle duality.
 
      Any quantum field can be decomposed into infinitely many Fourier modes. It can be shown that the dynamics of each Fourier mode is analogous to that of a simple harmonic oscillator. 
Hence, for each Fourier component there is a quantum zero point energy. This implies that the ground state energy (i.e. the lowest energy) of the quantum field, which is a sum of all these 
zero point energies, is infinite! This is not so much of a problem in non-gravitational quantum theory. The reason is as follows.
   
   The ground state of the total Hamiltonian encompassing all the quantum fields of nature is called the vacuum state (because, this state corresponds to `no particle' state, i.e. literally a 
vacuum like condition). A single particle state is analogous to the first excited state, and so on. The energy of a single particle state is also infinite (obviously, as the vacuum  or the 
lowest energy itself is infinite). Normally in experiments one deals with particle states, and hence, one can say that one will deal only with the difference of energy between the particle 
state and the vacuum. This is of course finite.
 So, by redefining the vacuum energy to be the `zero level' one can consistently talk about the observed finite energy of particles.  After all, the zero of a potential energy has no observable significance - one can add or subtract any constant quantity. It is only the change in the potential energy that is 
measurable. So, why should throwing away of vacuum energy be any different, right? Wrong.

The ground state actually has non-trivial significance. Firstly, vacuum state gets influenced when the boundary conditions are altered. For example, if one has two large, thin and parallel 
conducting plates, the electric field has to vanish on the plates. Because of this boundary condition, the Fourier modes corresponding to the electromagnetic (EM) field within the plates are 
discrete. That is, 
the EM wavelengths are quantized (just like it happens in the case of a stretched wire kept clamped at two points). The physicist Casimir showed that the difference between the vacuum energy 
outside the plates and between them is finite and increases with the plate separation. That means, if one takes two extremely smooth conducting plates, there should be a tiny force of attraction
 between them. This Casimir effect has been experimentally seen. Note that the force of attraction is not due  to EM force. Rather, it is because of the non-trivial vacuum or the ground state 
configuration of the vacuum EM field.  
  
   Secondly, throwing away the vacuum energy is strictly illegal, as all forms of energy warp the space-time geometry (i.e. even vacuum energy is a source of gravity). An infinite 
vacuum energy is of course a bigger worry when gravitational interaction is brought in. So, how does one include vacuum energy's contribution to gravity?

    The celebrated Russian physicist Zeldovich had drawn attention towards a very  interesting point. We know that the Minkowski metric tensor is invariant under Lorentz transformations and 
space-time translations. That is, for any observer in any arbitrary inertial frame, the metric tensor is $\eta_{\mu \nu}$. Also, the vacuum state must be invariant under Lorentz transformations 
as well as space-time translations.
 In other words, a `no particle' state in an inertial frame S must be a `no particle' state in any other inertial frame S$^\prime $. Otherwise, an observer in an inertial frame can find out her/his 
absolute velocity using a particle detector in vacuum, which is absurd. 

Therefore, the energy-momentum tensor of the vacuum must look the same in all inertial frames.
This means that the energy-momentum tensor corresponding to the vacuum 
state must be locally of the form $V \eta_{\mu \nu}$      (where  $V$  is a constant  vacuum energy density). Why? Because it  is the only 
second rank tensor invariant under both
 Lorentz transformations as well as space-time translations, so that vacuum or the `no particle' state in any inertial frame is still the`no particle' state in any other inertial frame.

  In a local inertial frame, the metric tensor is identical  to the Minkowski tensor. Hence, in an arbitrary coordinate system,
 vacuum energy-momentum tensor must be of the form   $V g_{\mu \nu}$. This implies that both vacuum energy-momentum tensor and the  cosmological constant term have similar forms.
     Einstein's cosmological constant  term appears on the L.H.S. of Einstein equations in GR while the vacuum energy-momentum tensor sits on the right, along with the matter energy-momentum 
tensor.

 There arises an interesting possibility. Suppose,  Einstein's cosmological constant  $\Lambda $   and  the vacuum energy density $V$    are both infinite but their difference is finite and
 small. This 
tiny remnant can act as an effective cosmological constant. Now, that can explain all the current cosmological observations concerning  accelerated expansion of the universe.
   But there is a catch! Why should the two infinite quantities be so adjusted as to leave a tiny difference? This  is the so called cosmological constant fine tuning problem [9,10]. 

 The riddle of cosmological constant persists even though the original motivation for its introduction has long vanished after the discovery of Hubble's law. It is like the smile 
of the Chesire cat - the cat disappeared but its smile remained! Up there, Einstein would now be smiling with the kind of havoc that
 his `blunder' has been causing.

{\bf Figure caption}

Figure 1:

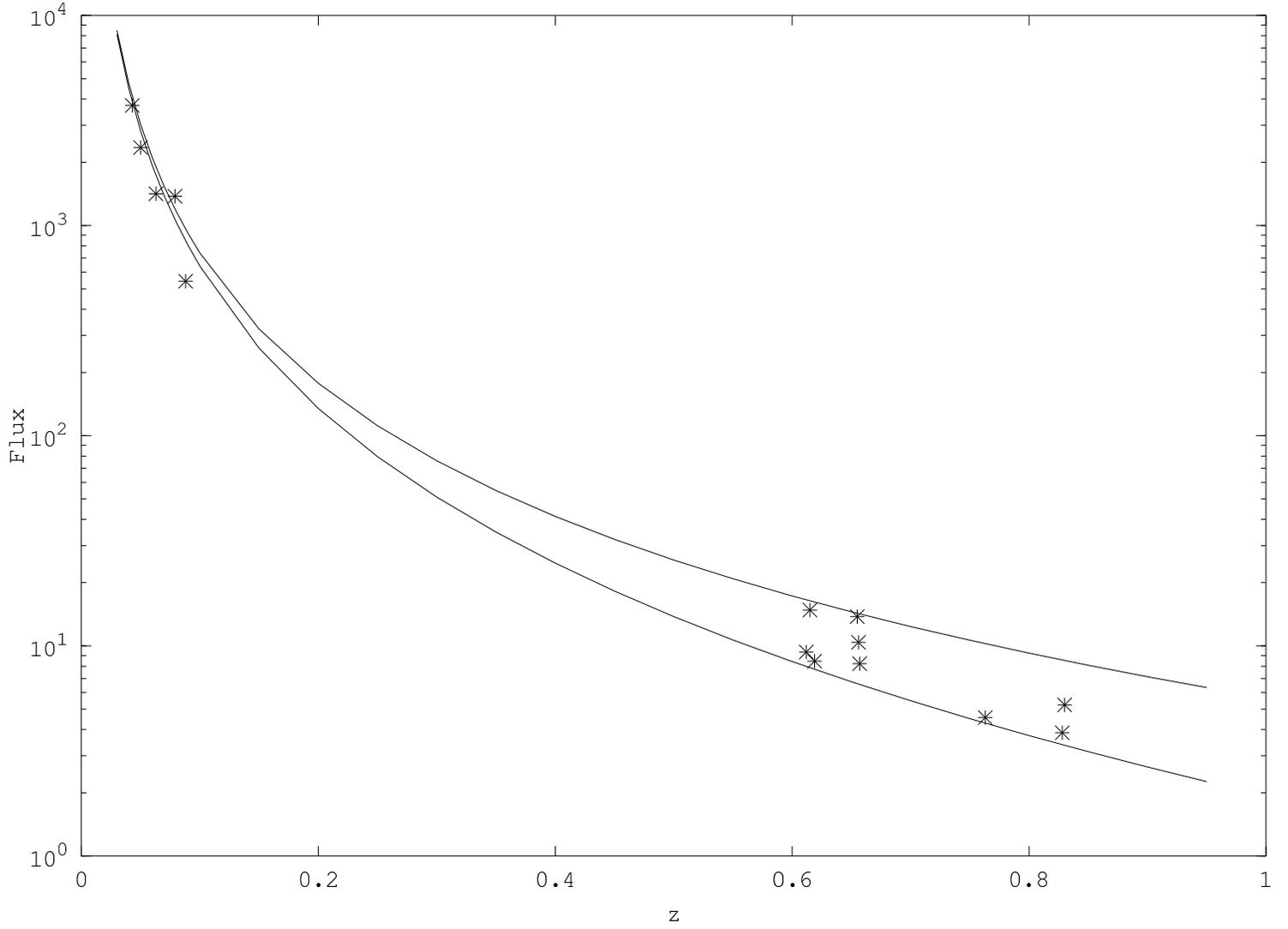
 In the above graph, each * symbol corresponds to a pair of measured redshift and peak flux, in units of $10^{-15}$ erg/s/cm$^2$, for 14 Type Ia supernovae (fluxes
 have been
 estimated from the effective peak magnitude $m^{eff}_B$ provided in ref [5]). The upper and lower solid lines represent the flux-redshift relations for a 
standard candle of luminosity $2 \times 10^{43}$ erg/s corresponding to luminosity-distances $d^{(M)}_L(z)$ and $d^{(\Lambda)}_L(z)$, respectively.

{\bf References}

[1] Perlmutter, S., et al., Astrophys. J. {\bf 483}, 565 (1997) 

[2] Schmidt, B., et al., Astrophys. J. {\bf 507}, 46 (1998) 

[3] Perlmutter, S., et al., Nature {\bf 391}, 51 (1998)
 
[4] Reiss, A., et al., Astron. J. {\bf 116}, 1009 (1998) 

[5] Perlmutter, S., et al., Astrophys. J. {\bf 517}, 565 (1999)

[6] Das Gupta, P., arXiv:1107.3460v2 [physics.hist-ph] (for an elementary exposition)

[7] Perlmutter, S., Physics Today, {\bf 56}, 53 (2003) (for an excellent introduction)

[8] Das Gupta, P., in Geometry, Fields and Cosmology, eds. B. R. Iyer and C. V. Vishveshwara (Kluwer Academic Publishers, Dordrecht, 1997), 525
 (for an introduction at a graduate course level)
 
[9] Weinberg, S., 1989, Rev. Mod. Phys. {\bf 61}, 1 

[10] Miao Li, Xiao-Dong Li, Shuang Wang, Yi Wang, Comm. Theor. Phys. {\bf 56}, 525 (2011)  (for a recent comprehensive review)
\end{document}